\documentclass[journal,final]{IEEEtran}

\usepackage{cite}
\usepackage{graphicx}
\usepackage{amsmath,amssymb}
\usepackage[caption=false,font=footnotesize]{subfig}
\usepackage{url}
\usepackage{hyperref}
\usepackage{xcolor}
\usepackage{tikz}
\usepackage{bm}
\usepackage{array}
\usetikzlibrary{circuits.ee.IEC, positioning}
\tikzstyle{bus}=[rectangle, inner sep=0pt, minimum height=0pt, minimum width=30pt, draw]

\usepackage{algorithm}
\usepackage{algorithmic}

\newcommand{\tensor}[1]{\mathbf{#1}}
\newcommand{\mexp}[1]{\mathbb{E}\left\{{#1}\right\}}
\DeclareMathOperator*{\argmin}{arg\,min}
\newcommand{\PreserveBackslash}[1]{\let\temp=\\#1\let\\=\temp}
\newcolumntype{C}[1]{>{\PreserveBackslash\centering}p{#1}}

\usepackage{epstopdf}
\usepackage{tikzpowergrid}

\begin{document}

\title{PDF estimation for power grid systems via sparse regression}%
\author{Xiu Yang, %
  David~A.~Barajas-Solano, %
  W.~Steven Rosenthal, %
  Alexandre~M.~Tartakovsky}

\IEEEpubid{0000--0000/00\$00.00~\copyright~2016 IEEE}

\maketitle

\begin{abstract}
  We present a numerical approach for estimating the probability density function (PDF) of quantities of interest (QoIs) of power grid systems subject to uncertain power generation and load fluctuations.
  In our approach, generation and load fluctuations are modeled by means of autocorrelated-in-time random processes, which are approximated in terms of a finite set of random parameters by means of Karhunen-Lo\`{e}ve approximations.
  The map from random parameters to QoIs is approximated by means of Hermite
  polynomial expansions. We propose a new approach based on compressive sensing to estimate the
  coefficients in the Hermite expansions from a small number of realizations (sampling points). 
  Linear transforms identified by iterative rotations are introduced to improve
  the sparsity of the Hermite representations, exploiting the intrinsic
  low-dimensional structure of the map. As such, the proposed approach
  significantly reduces the required number of sampling points to achieve a
  given accuracy compared to the standard least squares method.
  The proposed approach is employed to estimate the PDF of relative angular velocities and bus voltages of systems of classical machines driven by autocorrelated random generation.
  More accurate PDF estimates---as measured by the Kullback-Leibler divergence---are achieved using fewer realizations than required by basic Monte Carlo sampling.
\end{abstract}


\section{Introduction}
\label{sec:intro}

\IEEEPARstart{M}{odern} design and operation of power grid systems require accounting for the various sources of uncertainty in power generation and demand.
Adopting a probabilistic framework, in this manuscript we propose a method to estimate the probability density function (PDF) of the power grid system states to quantify the impact of random variability in generation and demand on the uncertainty in (and safety and reliability of) power grid systems.
The literature on uncertainty quantification (UQ) provides various strategies for propagating uncertainty in generation and load through the differential-algebraic equations (DAEs) governing the transient state of power grid systems.
A possible approach is the so-called PDF method for stochastic differential equations driven by autocorrelated noise~\cite{barajas-solano-2016-probabilistic}.
The PDF method has been applied successfully to analyzing the small-signal and transient stability properties of power grid systems~\cite{wang-2015-probabilistic,barajas-solano-probabilistic}, but can be computationally demanding for transient analysis with many degrees of freedom.

In this manuscript we propose an alternative approach for estimating the distribution of the power grid system quantities of interest (QoI) based on iterative sparse regression~\cite{yang-2016-enhancing}.
Here, QoIs are the power grid states such as the generator angular velocities and bus voltages at a given time.
In our approach, power generation and load fluctuations are modeled as autocorrelated random processes of time, which we represent in terms of a finite set of independent and identically distributed (iid) Gaussian random variables via truncated Karhunen-Lo\`{e}ve approximations.
We employ Hermite polynomial expansions to approximate the map from random variables to a QoI, i.e., we construct a surrogate model for the QoI.
Then the PDF of QoI is estimated by sampling the surrogate model which is less expensive than running the full model.
Instead of employing Galerkin projection, we estimate the expansion coefficients in the surrogate model from realizations of the transient state by sparse regression, which requires much fewer realizations of the transient state of the full model than the standard least squares method.
By improving the efficiency of the sparse regression methodology, we can further reduce the number of realizations required to construct the surrogate model.
For this purpose, we iteratively compute a linear transformation in the space of random variables that transforms the original random variables to a new set of random variables
to improve the sparsity of the expansion coefficients. We then use the information based on the enhanced sparsity to reduce the dimension of the representation of the uncertainty.

We apply our proposed approach to analyzing two power grid systems: The WECC 3-generator, 9-bus system~\cite{anderson-2008-power}, and the 10-generator, 39-bus New England system ~\cite{Pai1989}.
For these two systems, we estimate the PDF of the angular velocity of synchronous generators relative to the swing generator.
For the New England system we also estimate the PDF of the bus voltages.
We compare PDF estimates computed by our approach with kernel density estimates computed directly from Monte Carlo (MC) realizations of the full model.
For this purpose, we compute the Kullback-Leibler (KL) divergence~\cite{Kullback51} of PDF estimates with respect to an accurate MC estimate of the density.
For all cases considered, the KL divergence of the sparse sampling PDF estimate is reduced by $75\%$ compared to the KL divergence of the kernel density estimate computed from the same number of MC realizations of the full model that were employed to compute the surrogate model.

\IEEEpubidadjcol 

\section{Problem formulation}
\label{sec:formulation}

We consider power grid systems driven by random-in-time mechanical power and load.
Let $(\Omega, \mathcal{F}, \mathbb{P})$ be a complete probability triple, where $\Omega$ is the outcome space, and $\mathbb{P}$ is the probability measure over the $\sigma$-algebra of events $\mathcal{F}$.
Accounting for the sources of uncertainty, the transient behavior of power grid systems over a time window $[0, t_{\max}]$ obeys a set of stochastic differential algebraic equations of the form
\begin{align}
  \label{eq:sdae-d}
  \dot{\mathbf{x}} &= \mathbf{f} \bm{(} \mathbf{x}, \mathbf{y}; \lambda(t), \tilde{\omega} \bm{)}, & x(0) &= x_0,\\
  \label{eq:sdae-a}
  \bm{0} &= \mathbf{g} \bm{(} \mathbf{x}, \mathbf{y}; \lambda(t), \tilde{\omega} \bm{)}, & y(0) &= y_0,
\end{align}
where $\mathbf{x}$ is the vector of transient states (and its controllers), $\mathbf{y}$ is the vector of algebraic states, and $\tilde{\omega} \in \Omega$ is an outcome.
Here, $\lambda$ denotes discrete events (e.g., faults).
We are interested in estimating the PDF $p_u(U)$ of a quantity of interest $u = u \bm{(} \mathbf{x}(t_{\max}; \tilde{\omega}), \mathbf{y}(t_{\max}; \tilde{\omega}) \bm{)} \in D$, $D \subseteq \mathbb{R}$, of the system's state.

Without loss of generality, we restrict our attention to power systems driven by uncertain power injection.
  We denote by $P^m_k(t; \tilde{\omega})$ the $k$th generator's mechanical power injection random process.
In order to account for the non-Gaussian character of renewable generation, we model each $Y_k = \ln P^m_k$ as a square-integrable, stationary Gaussian process.
Furthermore, we assume that the Gaussian processes $Y_k$ are uncorrelated with each other\footnote{Mutually correlated generation fluctuations can be considered via multi-correlated K-L expansions (e.g.~\cite{cho-2013-karhunen}).}, with covariance kernel $C_k(t, s) = \langle Y_k(t) Y_k(s) \rangle$, where $\langle \cdot \rangle$ denotes ensemble average.
We approximate each $Y_k$ by means of its Karhunen-Lo\`{e}ve (K-L) expansion truncated to $d_k$ terms,
\begin{equation}
  \label{eq:kl-Pm}
  Y_k(t; \tilde{\omega}) = \left \langle Y_k \right \rangle + \sum^{d_k}_{i = 1} \sqrt{\gamma^k_i} \phi^k_i(t) \xi^k_i(\tilde{\omega}),
\end{equation}
where $\{ \xi^k_i(\tilde{\omega}) \}^{d_k}_{i = 1}$ is a set of iid standard Gaussian random variables, and $\{ \gamma^k_i, \phi^k_i \}^{d_k}_{i = 1}$ is the set of K-L eigenvalue and eigenfunction pairs, satisfying the Fredholm integral equation of the second kind
\begin{equation*}
  \int^T_0 C_k(t, s) \phi^k_i(s) \, \mathrm{d} s = \gamma^k_i \phi^k_i(t).
\end{equation*}

We assemble all sets of random variables into the random vector
\begin{equation*}
  \bm{\xi} \equiv \left ( \{ \xi^1_i \}^{d_1}_{i = 1}, \dots, \{ \xi^n_i \}^{d_n}_{i = 1} \right )^{\top}
\end{equation*}
with standard multivariate normal joint density $\rho(\bm{\xi}) \equiv \mathcal{N}(\bm{\xi} | \bm{0}, \tensor I)$ and support $\Gamma \equiv \mathbb{R}^d$, where $d = \sum_i d_i$ is the so-called stochastic dimension of the truncated problem.


\section{Surrogate Model}
\label{sec:surrogate}

The SDAEs~\eqref{eq:sdae-d} and \eqref{eq:sdae-a} together with the truncated K-L expansion~\eqref{eq:kl-Pm} implicitly define a map $u(\bm{\xi}) : \Gamma \to D$ from each random input vector to a value of the QoI.
We approximate such maps using the truncated Hermite polynomial expansion
\cite{GhanemS91, XiuK02}:
\begin{equation}
  \label{eq:gpc}
  u(\bm\xi) \approx \tilde{u}(\bm\xi) \equiv \sum^N_{i = 1} c_i \psi_i(\bm\xi),
\end{equation}
where $\{ \psi_i(\bm\xi) \}^N_{i = 1}$ is the set of normalized Hermite
polynomials orthogonal with respect to $\rho(\bm\xi)$, i.e.,
\begin{equation}
  \label{eq:gpc-orthogonality}
  \mexp{\psi_i(\bm{\xi}) \psi_j(\bm{\xi})} \equiv \int_{\Gamma} \psi_i(\bm\xi) \psi_j(\bm\xi) \rho(\bm\xi) \, \mathrm{d} \bm\xi = \delta_{ij},
\end{equation}
where $\delta_{ij}$ is the Kronecker delta function, and $\mathbb{E} \{ \cdot \}$ denotes the expectation $\int_{\Gamma} (\cdot) \rho(\bm{\xi}) \, \mathrm{d} \bm{\xi}$.

The standard approach for approximating $p_u$ is via MC simulations, i.e., to
simulate a large ensemble of $M$ QoI samples $\{u^q\}_{q=1}^M \equiv
\{u(\bm\xi^q)\}_{q=1}^M$ based on the set of independent input samples
$\{\bm\xi^q\}_{q=1}^M$, and then use the kernel density estimate method
\cite{Parzen62} to approximate $p_u$.
Instead, we propose sampling the surrogate model $\tilde u$ of~\eqref{eq:gpc} to approximate $p_u$.
We note that sampling $\tilde u$ simply requires evaluating a polynomial, which is less costly than simulating the original SDAE system \eqref{eq:sdae-d}--\eqref{eq:sdae-a}.
As such, this surrogate based method is useful when sampling $u$ is expensive and many samples are needed.
For example, in some optimization problems, the objective function or constraints require the computation of the PDF or statistics (e.g., mean, variance) of $u$ (e.g, \cite{Dentcheva06,Krokhmal11}).
Another example is Bayesian inference for identifying the model parameters, in which a large number of samples of $u$ (typically $\mathcal{O}(10^5)$) is needed \cite{MarzoukNR07, LeiYLK17}.

The multivariate Hermite polynomials $\psi_i$ are constructed as the tensor product
of univariate Hermite polynomials. For a multi-index 
$\bm\alpha=(\alpha_1,\alpha_2,\cdots,\alpha_d)$, $\alpha_i\in\mathbb{N}\cup\{0\}$, we set
\begin{equation*}
  \psi_{\bm\alpha}(\bm\xi) =
  \psi_{\alpha_1}(\xi_1)\psi_{\alpha_2}(\xi_2)\cdots\psi_{\alpha_d}(\xi_d).
\end{equation*}
For example,
\[\psi_{1,0,1}=\psi_1(\xi_1)\psi_0(\xi_2)\psi_1(\xi_3)=\xi_1\xi_3.\]
For simplicity, we denote $\psi_{\bm\alpha_i}(\bm\xi)$ as $\psi_i(\bm\xi)$.
Based on this construction, the orthogonality in~\eqref{eq:gpc-orthogonality} holds, since for two different multi-indices 
$\bm\alpha_i=((\alpha_i)_{_1}, (\alpha_i)_{_2}, \cdots, (\alpha_i)_{_d})$ and 
$\bm\alpha_j=((\alpha_j)_{_1}, (\alpha_j)_{_2}, \cdots, (\alpha_j)_{_d})$, we
have 
\begin{multline*}
  \mexp{\psi_{i}(\bm\xi) \psi_{j}(\bm\xi)} = \delta_{\bm\alpha_i\bm\alpha_j}\\
  = \delta_{(\alpha_i)_{_1}(\alpha_j)_{_1}} \delta_{(\alpha_i)_{_2}(\alpha_j)_{_2}}\cdots \delta_{(\alpha_i)_{_d}(\alpha_j)_{_d}}.
\end{multline*}
For the expansion of $\tilde u$ with polynomials up to $P$th order, $|\bm\alpha|=\sum_{i=1}^d \alpha_i\leq P$ and $N=\binom{P+d}{P}$.


\section{Sparse Regression}
\label{sec:sparse-reg}

Given the sets of input and QoI samples, $\{\bm\xi^q\}_{q=1}^M$ and $\{u^q \equiv u(\bm{\xi}^q)\}_{q=1}^M$, respectively, constructing $\tilde u$ requires identifying the coefficients $\{c_i\}_{i=1}^N$ in Eq.~\eqref{eq:gpc}.
This is done by solving the linear system
\begin{equation}
  \label{eq:gpc-approx}
  \tensor \Psi \mathbf{c} \approx \mathbf{u}, 
\end{equation}
where $\mathbf{c} = (c_1, c_2, \cdots, c_N)^{\top}$, $\mathbf{u}=(u^1, u^2, \cdots, u^M)$, and $\tensor \Psi$ is the so-called measurement matrix with components $\Psi_{ij} = \psi_j(\bm\xi^i)$.
For $M > N$, the linear system~\eqref{eq:gpc-approx} is overdetermined, and the standard approach to approximate $\mathbf{c}$ is ordinary least squares fitting.
For $M < N$, the system is underdetermined, and a QR decomposition of $\tensor\Psi$ should be used before applying ordinary least squares fitting~\cite{underdetermined-systems}.

Given sufficient regularity of $u$, the difference between $u$ and $\tilde u$ becomes smaller as more polynomials are included in the expansion of $\tilde u$; i.e., the error decreases with increasing $N$.
Often, the number of available samples of $u$ is smaller than the $N$ required to obtain a solution with the desired error, i.e., $M<N$ or even $M \ll N$.
Although ordinary least squares fitting with QR decomposition can be used in this case, a more accurate approach here is compressive sensing with $\ell_1$ minimization~\cite{CandesRT06, DonohoET06}:
\begin{equation}
  \label{eq:l1-minimization}
  (P_{1,\epsilon})\quad  \argmin_{\hat{\mathbf{c}}}\Vert \hat{\mathbf{c}}
  \Vert_1, \text{ subject to } \Vert \tensor\Psi \hat{\mathbf{c}} -\mathbf{u} \Vert_2\leq\epsilon,
\end{equation}
where $\Vert \cdot \Vert_p$ denotes the $\ell_p$ norm and $\epsilon$ is an estimate of the truncation error.
Theoretical analysis in \cite{CandesRT06, DonohoET06}
demonstrates that for $\tensor\Psi$ satisfying the restricted isometry condition,
the $\ell_1$ minimization yields an accurate estimate of a \emph{sparse} $\mathbf{c}$, where \emph{sparse} means that most $c_n$ coefficients are close to 0 and can be disregarded.
Intuitively, the restricted isometry property indicates that $\tensor \Psi$ is nearly orthonormal, i.e., that $\tensor\Psi^{\top} \tensor\Psi$ is close to the identity matrix.
We refer interested readers to \cite{CandesRT06} for more details.
A modified version of $\ell_1$ minimization, named \emph{reweighted $\ell_1$ minimization}, was proposed to improve the accuracy of approximating $\mathbf{c}$ \cite{CandesWB08, YangK13}:
\begin{equation}
    \label{eq:l1-minimization-reweighted}
    (P_{1,\epsilon}^W) \quad \argmin_{\hat{\mathbf{c}}}\Vert\tensor W
    \hat{\mathbf{c}} \Vert_1, \text{ subject to } \Vert \tensor\Psi
    \hat{\mathbf{c}} - \mathbf{u} \Vert_2\leq\epsilon,
\end{equation}
where $\tensor W \equiv \text{diag}(w_1, w_2, \cdots, w_d)$.
This minimization is performed iteratively: first, $(P_{1,\epsilon})$ is solved to obtain an initial guess $\hat{\mathbf{c}}^{(0)}$; then, we set $w^{(1)}_i = 1/(|\hat c_i^{(0)}|+\gamma)$ and $(P_{1,\epsilon}^W)$ is solved to obtain $\hat{\mathbf{c}}^{(1)}$.
These steps are repeated until convergence is achieved.  
Usually, only two to three iterations are performed, as more iterations don't provide significant improvement~\cite{CandesWB08, YangK13}.

In this work we propose employing the iterative rotations method developed in~\cite{LeiYZLB15, yang-2016-enhancing} to enhance the sparsity of $\mathbf{c}$.
This method aims to identify an orthonormal rotation matrix $\tensor A$ (i.e., satisfying $\tensor A \tensor A^{\top}=\tensor I$) that maps $\bm\xi$ to a new set of random variables $\bm\eta \equiv \tensor A \bm\xi$, where $\rho(\bm\eta)=\mathcal{N}(\bm\eta|\bm 0, \tensor I)$ due to the orthonormality of $\tensor A$.
In terms of $\bm{\eta}$, $\tilde u$ in~\eqref{eq:gpc} can be rewritten as
\begin{equation}
  \label{eq:alg_transform}
  \tilde u(\bm\xi) = \sum_{n=1}^Nc_n\psi_n(\bm\xi)=\sum_{n=1}^N \tilde c_n\psi_n(\tensor A\bm\xi) = \sum_{n=1}^N\tilde c_n\tilde{\psi}_n(\bm\eta).
\end{equation}
By enhancing the sparsity of $\tilde{\mathbf{c}}=(\tilde c_1, \tilde c_2, \cdots, \tilde c_N)$ with respect to that of $\mathbf{c}$, the accuracy of approximating the Hermite expansion coefficients by the solution of $(P_{1,\epsilon})$ (or $(P^W_{1,\epsilon})$) is expected to increase.

The rotation matrix $\tensor A$ can be found iteratively~\cite{yang-2016-enhancing} using the eigendecomposition of the gradient variance~\cite{Russi10,ConstantineDW14}
\begin{equation}
  \label{eq:grad_mat}
  \tensor G \equiv
  \mexp{\nabla u(\bm\xi) \nabla u(\bm\xi)^{\top}} = \tensor U\tensor \Lambda\tensor U^{\top}, \quad \tensor U\tensor U^{\top} = \tensor I,
\end{equation}
and setting $\tensor A=\tensor U^{\top}$. The matrix $\tensor U$ consists of columns of eigenvectors, and $\tensor \Lambda$ is a diagonal matrix of eigenvalues $\{\lambda_i\}_{i=1}^d$ with $\lambda_1\geq\lambda_2\cdots\geq\lambda_d\geq 0$.
The rotation $\bm\eta=\tensor A\bm\xi$ projects $\bm\xi$ onto the eigenvectors $\mathbf{U}_{(i)}$.
Consequently,
when the sequence $\{ \lambda_i \}$ decays rapidly,
$u$ primarily depends on the first few new random variables $\eta_i$.
That is, most of the variation of $u$ is concentrated along the directions of the corresponding eigenvectors.
Since $u$ is not known, $\tensor G$ is approximated in terms of $\tilde u$, i.e.,
\begin{equation}
  \label{eq:grad_mat_approx}
  \tensor G \approx
  \mexp{\nabla\tilde u(\bm\xi) \nabla \tilde u(\bm\xi)^{\top}},
\end{equation}
or,
\begin{equation}
  \label{eq:grad}
  \begin{aligned}
    G_{ij} & \approx\mexp{ \dfrac{\partial}{\partial\xi_i}\left(\sum_{n=1}^N c_n\psi_n(\bm\xi)\right) \dfrac{\partial}{\partial\xi_j}\left(\sum_{n'=1}^Nc_{n'}\psi_{n'}(\bm\xi)\right)} \\
    & = \mexp{\left(\sum_{n=1}^Nc_n\dfrac{\partial\psi_n(\bm\xi)}{\partial\xi_i}\right) \left(\sum_{n'=1}^Nc_{n'}\dfrac{\partial\psi_{n'}(\bm\xi)}{\partial\xi_j}\right)} \\
    & = \sum_{n=1}^N\sum_{n'=1}^Nc_nc_{n'} \mexp{\dfrac{\partial\psi_n(\bm\xi)}{\partial\xi_i}
      \dfrac{\partial\psi_{n'}(\bm\xi)}{\partial\xi_j}} \\
    & = \mathbf{c}^{\top} \tensor K_{ij} \mathbf{c},
  \end{aligned}
\end{equation}
where $\tensor K_{ij}$ are matrices with components
\begin{equation}\label{eq:kernel}
  \begin{aligned}
    (K_{ij})_{kl}& = \mexp{\dfrac{\partial\psi_k(\bm\xi)}{\partial\xi_i}
                \dfrac{\partial\psi_{l}(\bm\xi)}{\partial\xi_j}} \\
& = \sqrt{(\alpha_k)_{_i}(\alpha_l)_{_j}} \delta_{(\alpha_k)_{_i}-1(\alpha_l)_{_i}}
    \delta_{(\alpha_k)_{_j}(\alpha_l)_{_j}-1}\\
& \prod_{\substack{m=1\\ m\neq i,m\neq j}}\delta_{(\alpha_k)_{_m}(\alpha_l)_{_m}},
\end{aligned}
\end{equation}
and the index $k$ in $\psi_k$ is the multi-index $\bm\alpha_k=((\alpha_k)_1,(\alpha_k)_2,\cdots, (\alpha_k)_d)$.
In~\eqref{eq:kernel}, the following property of univariate normalized Hermite polynomials is used:
\begin{equation}
\psi'_n(x)=\sqrt{n}\psi_{n-1}(x),\quad n\in\mathbb{N}\cup\{0\}, \quad \psi_{-1}(x)=0.
\end{equation}
Note that $\tensor G$ is a symmetric $d\times d$ matrix ($d$ is the number of random variables in the system) and only $d(d+1)/2$ of its entries need to be computed.
\begin{algorithm}
  \caption{Sparsity-enhancing $\ell_1$ minimization with iterative rotations.}
  \label{algo:cs_rot}
  \begin{algorithmic}[1]
    \STATE Generate input samples $\{ \bm\xi^q \}^M_{q = 1}$ from the distribution $\rho(\bm\xi)$.%
    \STATE Generate QoI samples $\{ u^q \equiv u(\bm\xi^q) \}^M_{q = 1}$ by simulating SDAE system~\eqref{eq:sdae-d}--\eqref{eq:sdae-a} with K-L expansions \eqref{eq:kl-Pm}.%
    \STATE Construct the measurement matrix $\tensor \Psi$ by setting $\Psi_{ij}=\psi_j(\bm\xi^i)$.%
    \STATE Solve the optimization problem $(P_{1,\epsilon})$ \eqref{eq:l1-minimization} to compute $\hat{\mathbf{c}}$.
    If the reweighted $\ell_1$ method is employed, solve $(P_{1,\epsilon}^W)$ \eqref{eq:l1-minimization-reweighted} instead.%
    \STATE Set $l=0$, $\eta^{(0)}=\bm\xi$, $\tilde{\mathbf{c}}^{(0)}=\hat{\mathbf{c}}$.%
    \STATE Construct $\tensor G^{(l+1)}$ from $\hat{\mathbf{c}}^{(l)}$ using~\eqref{eq:grad}, then compute the eigendecomposition $\tensor G^{(l+1)}=\tensor U^{(l+1)}\tensor\Lambda^{(l+1)}(\tensor U^{(l+1)})^{\top}$.
    \STATE Define $\bm\eta^{(l+1)}=(\tensor U^{(l+1)})^{\top}\bm\eta^{(l)}$, and compute samples $(\bm\eta^{(l+1)})^q=(\tensor U^{(l+1)})^{\top}(\bm\eta^{(l)})^q$, $q = 1, 2, \dots, M$.%
    \STATE Update the measurement matrix $\tensor\Psi^{(l+1)}$ with $\Psi^{(l+1)}_{ij}=\tilde{\psi}_j \bm{(} (\bm\eta^{(l+1)})^i \bm{)}$.%
    \STATE Solve the optimization problem $(P_{1,\epsilon^{(l+1)}})$ and set $\tilde{\mathbf{c}}^{(l+1)}=\hat{\mathbf{c}}$. If the reweighted $\ell_1$ method is employed, solve $(P_{1,\epsilon^{(l+1)}}^W)$ instead.
    \STATE Set $l=l+1$.
    If $l=l_{\max}$, stop; otherwise, go to Step 6.
  \end{algorithmic}
\end{algorithm}

The iterative rotation algorithm proposed in
\cite{yang-2016-enhancing} is summarized as Algorithm~\ref{algo:cs_rot}.
At each iteration, $\tilde{\mathbf{c}}$ is used from the previous step to compute $\tensor G$ based on Eq.~\eqref{eq:grad}, and its eigendecomposition (Step 6).
$\bm\eta^{(l)}$ is updated from $\bm\eta^{(l+1)}$ as $\bm\eta^{(l+1)}=(\tensor U^{(l+1)})^{\top}\bm\eta^{(l)}$ (Step 7).
Once the maximum number of iterations $l_{\mathrm{max}}$ is reached, $\mathbf{A}$ is set as
\begin{equation*}
  \tensor A = (\tensor U^{(1)} \tensor U^{(2)} \cdots \tensor U^{(l_{\max})})^{\top}.
\end{equation*}
The maximum iteration numbers $l_{\max}$ is usually set at two to three according to the authors' experience since more iterations will not improve the accuracy significantly.
A more sophisticated stopping criterion can be designed by measuring the distance between $\tensor U^{(l)}$ and the identity matrix or permutation matrix.
More details can be found in \cite{yang-2016-enhancing}.
In the present work, we set $l_{\max}=2$.
The $\ell_1$ minimization problems are solved using the MATLAB package \texttt{SPGL1}~\cite{BergF08, spgl1}.
In practice, the thresholds $\epsilon$ and $\epsilon^{(l)}$ are estimated by cross-validation since these thresholds
are not known \emph{a priori}. One such technique for estimating the threshold
(based on \cite{DoostanO11}) is summarized in Algorithm~\ref{algo:cross}.
\begin{algorithm}
\caption{Cross-validation for estimating the error $\epsilon$.}
\label{algo:cross}
\begin{algorithmic}[1]
  \STATE Divide the $M$ output samples into $M_r$ reconstruction ($\bm u_r$) and $M_v$ validation ($\bm u_v$) samples, and divide the measurement matrix $\tensor\Psi$ correspondingly into $\tensor\Psi_r$ and $\tensor\Psi_v$. \\[-1pc]
  \STATE Choose multiple values for $\epsilon_r$ such that the exact error $\Vert\tensor\Psi_r\bm c-\bm u_r\Vert_2$ of the reconstruction samples is within the range of $\epsilon_r$ values.
  \STATE For each $\epsilon_r$, solve $(P_{h,\epsilon})$ with $\bm u_r$ and $\tensor\Psi_r$ to obtain $\hat{\bm c}$, then compute $\epsilon_v=\Vert\tensor\Psi_v\hat{\bm c}-\bm u_v\Vert_2$.
  \STATE Find the minimum value of $\epsilon_v$ and its corresponding $\epsilon_r$. Set $\epsilon=\sqrt{M/M_r}\epsilon_r$.
\end{algorithmic}
\end{algorithm}

The iterative rotation procedure described above can be exploited to reduce the stochastic dimension of the problem.
The size of $\bm\eta$ can be reduced according to the magnitude of the eigenvalues $\{ \lambda^{(l_{\mathrm{max}})}_i \}$ by setting a threshold $\theta$ and truncating the sequence after $d^\ast$ such that $\sum_{k=1}^{d^\ast} \lambda^{(l_{\mathrm{max}})}_k > \theta$.
Similar methods for model reduction in random space have been proposed in the literature (e.g., active subspaces~\cite{ConstantineDW14}, basis adaptation~\cite{TipiG14}).
However, in this work the rotation matrix is computed in a different manner than existing methods.
It is designed specifically for limited data problems, and it takes advantage of an accurate surrogate model of $u$ based on sparse regression. 
As such, in this scenario, the proposed method provides more accurate guidance for dimension reduction based on this accurate surrogate model.
After truncating the random variables $\bm\eta$ to $\bm\eta^\ast=(\eta^{(l_{\mathrm{max}})}_1, \dots, \eta^{(l_{\mathrm{max}})}_{d^\ast})^{\top}$ , the polynomial order can be raised to $P$ to $P^\ast$, $P<P^\ast$, to better describe the variance of $u$.
Then, $(P_{1,\epsilon})$ (or $(P_{1,\epsilon}^W)$) is solved with a new matrix $\tensor \Psi^\ast$ with components $\Psi^\ast_{ij}=\psi^\ast_j((\bm\eta^\ast)^i)$ to obtain $\mathbf{c}^\ast$, resulting in the approximation
\begin{equation}
  u(\bm \xi) \approx u^\ast(\bm \eta^\ast) = \sum_{i = 1}^{N^\ast} c^\ast_i \psi^\ast_i(\bm \eta^\ast),
\end{equation}
where $N^\ast = \binom{P^\ast + d^\ast}{d^\ast}$ and $N^\ast < N$.
This procedure is summarized in Algorithm \ref{algo:cs_trunc}.

\begin{algorithm}
  \caption{$\ell_1$ minimization with dimension reduction after iterative rotations.}
  \label{algo:cs_trunc}
  \begin{algorithmic}[1]
    \STATE Run Algorithm \ref{algo:cs_rot}.
    \STATE Decide the truncation dimension $d^\ast$ based on the eigenvalues $\{ \lambda^{(l_{\mathrm{max}})}_i \}$.
    For example, set $\sum_{i=1}^{d^\ast} \lambda^{(l_{\mathrm{max}})}_i>0.95\sum_{i=1}^{d} \lambda^{(l_{\mathrm{max}})}_i$.
    \STATE Introduce new random variables $\bm\eta^\ast=(\eta^{(l_{\mathrm{max}})}_1, \cdots, \eta^{(l_\mathrm{max})}_{d^\ast})^{\top}$, then compute samples $(\bm\eta^\ast)^q$ based on samples $(\bm\eta^{(l_{\mathrm{max}})})^q$, $q = 1, 2, \dots, M$.
    \STATE Compute the measurement matrix $\tensor\Psi^\ast$ with $\Psi^\ast_{ij}=\psi^\ast_j \bm{(} (\bm\eta^\ast)^i \bm{)}$.
    \STATE Solve the optimization problem $(P_{1,\epsilon^\ast})$ and set $\mathbf{c}^\ast=\hat{\mathbf{c}}$. If the reweighted $\ell_1$ method is employed, solve $(P_{1,\epsilon^\ast}^W)$ instead. 
  \end{algorithmic}
\end{algorithm}

Note that unlike the difference between $u$ and $\tilde u$, for which computing the $L_2$ error of the approximation is possible,
the accuracy of approximating $u$ by $u^\ast$ cannot be evaluated in the same manner, as $\bm\xi$ and $\bm\eta^\ast$ are defined on different spaces of different dimension.
Instead, the various proposed approximations are evaluated by comparing the PDFs of $u$, $\tilde{u}$, and $u^\ast$. 
These PDFs are approximated via kernel density estimation from a large number of samples of the corresponding surrogate models.

\section{Numerical experiments}
\label{sec:experiments}

In this section, the sparse regression approach is employed to estimate the PDF $p_u(U)$ of various QoIs $u$ in a power grid system. 
Specifically considered are systems of synchronous machines driven by uncertain mechanical power injections.
Surrogate models $\tilde{u}$ (iterative rotation without stochastic dimension reduction) and $u^\ast$ (with stochastic dimension reduction) are constructed from $M$ MC samples of the SDAE system~\eqref{eq:sdae-d}--\eqref{eq:sdae-a} with K-L expansions~\eqref{eq:kl-Pm}. 
Each surrogate is sampled $10^4$ times, and the corresponding PDF estimate is calculated employing kernel density estimation~\cite{Parzen62}.

To evaluate the accuracy of the PDF estimation by the sparse regression approach, a reference PDF is computed from $M_{\mathrm{ref}} = 10^4 \gg M$ MC samples. 
The PDFs are compared by computing the Kullback-Leibler (KL) divergence,
\begin{equation}
  K\!L(P \Vert Q) = \int_D p(x) \log \frac{p(x)}{q(x)} \, \mathrm{d} x,
\end{equation}
which measures the difference between two distributions (or densities) $P$ and $Q$.
To evaluate the efficiency of the sparse regression approach, PDFs estimated by the sparse regression approach are also compared with kernel density estimates computed directly from the $M$ MC samples of the full model used to construct the surrogate models.
The bandwidth $h$ used for kernel density estimation is taken as~\cite{silverman1986density}
\begin{equation*}
  h=1.06\hat\sigma n^{-1/5},
\end{equation*}
where $n$ is the number of samples and $\hat\sigma$ is the sample standard deviation.

\subsection{WECC 3-generator, 9-bus system}
\label{sec:case9}

\tpgset{BusConnectionSpacing}{.25cm}	
\tpgset{BusLabelOffset}{0ex}			
\tpgset{BusLength}{30pt}				
\tpgset{BusWidth}{2pt}				
\tpgset{LineTurnOffset}{0.5}			
\tpgset{LoadArrowLength}{1}			
\tpgset{GeneratorOffset}{.5}			
\tpgset{InputPowerArrowLength}{1}		

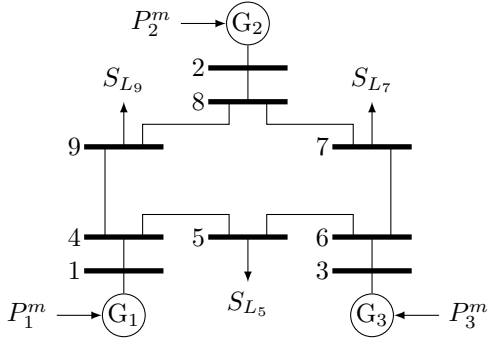
\begin{figure}[!t]
  \centering
  \begin{tikzpicture}[x = 0.6cm, y = 0.6cm]
    \bus(4)(0,0);%
    \bus(8)(2.75,3);%
    \bus(6)(5.5,0);%
    \bus(9)(0,2);%
    \bus(7)(5.5,2);%
    \bus(5)(2.75,0);%
    \bus(1)(0,-.75);%
    \bus(2)(2.75,3.75);%
    \bus(3)(5.5,-.75);%
    \lin(1,0)(4,0);%
    \lin(3,0)(6,0);%
    \lin(4,-1)(9,-1);%
    \lin(4,1)(5,-1);%
    \lin(5,1)(6,-1);%
    \lin(6,1)(7,1);%
    \lin(9,1)(8,-1);%
    \lin(7,-1)(8,1);%
    \lin(8,0)(2,0);%
    \gen(1)(1,-1);%
    \gen(2)(2,1);%
    \gen(3)(3,-1);%
    \load[$S_{L_9}$](9,0)(1);%
    \load[$S_{L_7}$](7,0)(1);%
    \load[$S_{L_5}$](5,0)(-1);%
    \powin(1)(west,1);%
    \powin(2)(west,1);%
    \powin(3)(east,1);%
  \end{tikzpicture}
  \caption{Schematic of the WECC 3-generator, 9 buses power system~\cite{anderson-2008-power}.}
  \label{fig:case9-schematic}
\end{figure}

The WECC 3-generator, 9-bus system~\cite{anderson-2008-power}, shown in Fig.~\ref{fig:case9-schematic} is considered.
The system consists of 3 classical synchronous generators, each driven by lognormally distributed mechanical power injections. as described in Section~\ref{sec:formulation}.
For classical generators, Eqs.~\eqref{eq:sdae-d} and \eqref{eq:sdae-a} can be rewritten as
\begin{align*}
  \dot{\delta}_k &= \omega_{\mathrm{B}} (\omega_k - \omega_{\mathrm{s}}),\\
  2 H_k \dot{\omega}_k &= - D_k (\omega_k - \omega_{\mathrm{s}}) - P^e_k(\bm{\delta}) + P^m_k(t; \tilde{\omega}),\\
  P^e_k(\bm{\delta}) &= \sum^n_{i = 1} E_k E_i \left ( G_{ki} \cos (\delta_k - \delta_i) + B_{ki} \sin (\delta_k - \delta_i) \right ),
\end{align*}
for $k \in [1, n]$, where $n$ is the number of generators.
Here, $\omega_k$ is the angular velocity [$\mathrm{rad}$ $\mathrm{s}^{-1}$] of the $k$th machine, $\bm{\delta} \equiv (\delta_1, \dots, \delta_n)^{\top}$ is the vector of generator phase angles [$\mathrm{rad}$], $H_k$ [$\mathrm{s}$] and $D_k$ [p.u.] are the generator's inertia and damping constants, respectively, $\omega_{\mathrm{s}}$ is the synchronous velocity [$\mathrm{rad}$ $\mathrm{s}^{-1}$], $\omega_{\mathrm{B}}$ is the base velocity [$\mathrm{rad}$ $\mathrm{s}^{-1}$], $G_{ij}$ and $B_{ij}$, $i, j \in [1, n]$, are the transfer conductances and susceptances, respectively [p.u.], $P^e_k$ is the active generated power [p.u.], and $P^m_k$ [p.u.] is the mechanical power injection [p.u.].

The mean mechanical power injections are $\langle P^{\mathrm{m}}_1 \rangle = 0.7128\text{ [p.u.]}$, $\langle P^{\mathrm{m}}_2 \rangle = 2.00\text{ [p.u.]}$, and $\langle P^{\mathrm{m}}_3 \rangle = 0.48\text{ [p.u.]}$, with equal standard deviation $\sigma = 0.05\text{ [p.u.]}$. For all generators, the dynamics of the mechanical power injections are modeled by K-L expansions, truncated at $25$ terms, of the exponential covariance kernel $C_k(t, s) = \sigma^2_{Y_k} \exp \{ -|t - s| / \lambda \}$, with correlation length $1.8 \text{ s}$. The stochastic dimension of the truncated problem is therefore $75$.

Starting from deterministic equilibrium initial conditions, the system is simulated for $t_{\max} = 10\text{ s}$.
At $1\text{ s}$, the system is subjected to a self-clearing 3-phase fault at the terminal of generator 2 with duration $0.8 \mathrm{CTT}$, where $\mathrm{CTT} = 0.189\text{ s}$ is the critical clearing time for the same fault and power system, starting from equilibrium initial conditions, but with no power injection uncertainty.

\begin{figure}[h]
  \centering
  \includegraphics[width=0.36\textwidth]{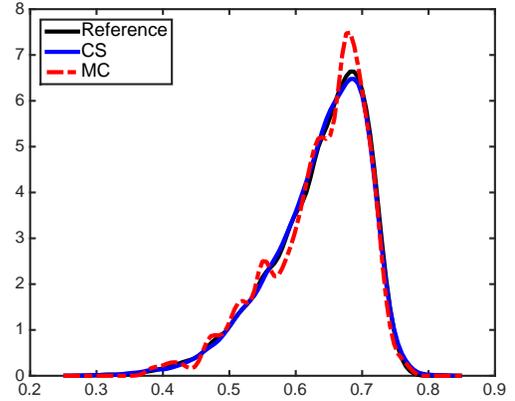}
  \caption{WECC 3-generator, 9-bus system: Comparison of the PDF of the relative angular velocity $\omega^\ast$ at time $t_{\mathrm{max}}$. ``MC'' indicate kernel density estimate computed from $500$ MC simulations; ``CS'' indicates kernel density estimate based on $10^4$ samples of $u^\ast$, which is constructed from the $500$ MC simulations; ``Reference'' indicates the reference kernel density estimate computed from $10^4$ MC simulations.}
  \label{fig:ex1}
\end{figure}

The aim is estimate the PDF of the angular velocity of generator 2 with respect to the swing generator 1 at time $t=t_{max}$, $u = \omega^\ast(t_{\max}) = \omega_2(t_{\max})-\omega_1(t_{\max})$.
$M = 500$ MC samples of the full system are used to construct the surrogate models $\tilde{u}$ and $u^\ast$ by the numerical method in Section~\ref{sec:sparse-reg}.
Then each surrogate model is sampled $10^4$ times to estimate the PDF of the QoI $u$.
First, Algorithm~\ref{algo:cs_rot} was run with $P = 2$ to obtain $\tilde u$.
Next, based on eigenvalues $\{ \lambda^{(l_{\mathrm{max}})}_i \}$ and the rotation $\tensor A$, and the parameter choices $d^\ast = 10$ and $P^\ast = 4$, the model for $u^\ast$ is constructed.
A reference kernel density estimate of the PDF is computed directly from $10^4$ samples of the full system.
Finally, for comparison, a kernel density estimate is directly computed from the $M = 500$ samples employed to construct the surrogate models.
Both MC PDF estimates, as well as the PDF estimate from $u^\ast$, are presented in Fig.~\ref{fig:ex1}.

KL divergences of the PDF estimates with respect to the reference PDF are presented in Table~\ref{tab:ex1}, which provides a quantitative understanding of Fig.~\ref{fig:ex1}.
PDFs were constructed, and their KL divergences computed, for $50$ independent sample sets, with each sample set consisting of $500$ samples. 
It can be seen that direct PDF estimation from $500$ MC samples is less accurate than PDF estimation by the sparse regression approach.
Similarly, it can be seen that the difference between PDF estimates from the $\tilde u$ (rotation without truncation) and $u^\ast$ (truncation after rotation) models is small.
\begin{table}
  \centering
  \caption{WECC 3-generator, 9-bus system: Mean KL divergence of various approximations to the PDF of the relative angular velocity $\omega^\ast$ at time $t_{\mathrm{max}}$, computed using different methods (with $500$ samples), with respect to the reference density.}
  \begin{tabular}{C{8em}C{8em}C{8em}}
    \hline\hline
      MC & $\tilde u$ & $u^\ast$ \\
    \hline
     $0.0235$  & $0.0081$  & $0.0056$ \\
    \hline\hline
  \end{tabular}
  \label{tab:ex1}
\end{table}

\begin{figure}[h]
  \centering
  \includegraphics[width=0.36\textwidth]{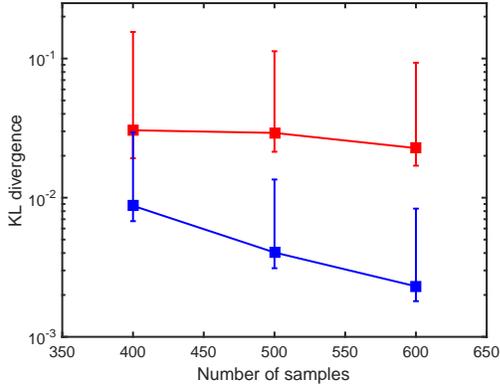}
  \caption{WECC 3-generator, 9-bus system: Comparison of mean KL divergence of approximations to the PDF of $\omega^\ast$ at time $t_{\mathrm{max}}$ with respect to the reference density.
    Red bars indicate direct kernel density estimate of PDF from MC simulations; blue bars indicate kernel density estimate of PDF from samples of $u^\ast$.}
  \label{fig:ex1_kl}
\end{figure}

The accuracy of the sparse regression method is examined for amounts of available data, $M$, varying from $400$ to $600$.
For each dataset size, $50$ independent datasets are examined, surrogate models are constructed, and KL divergences computed.
Figure~\ref{fig:ex1_kl} shows the mean KL divergence together with $99\%$ confidence intervals indicated by error bars.
The sparse regression method outperforms MC estimation for the same dataset size.

\subsection{10-generator New England system}
\label{sec:case39}

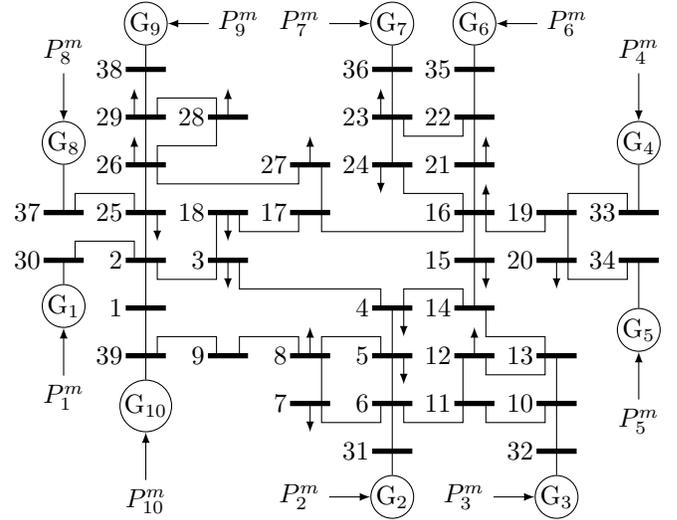
\begin{figure}[!t]
\centering
\begin{tikzpicture}[x = .43in, y = .25in]
	\bus(1)(0,2);
	\bus(2)(0,3);
	\bus(3)(1,3);
	\bus(4)(3,2);
	\bus(5)(3,1);
	\bus(6)(3,0);
	\bus(7)(2,0);
	\bus(8)(2,1);
	\bus(9)(1,1);
	\bus(10)(5,0);
	\bus(11)(4,0);
	\bus(12)(4,1);
	\bus(13)(5,1);
	\bus(14)(4,2);
	\bus(15)(4,3);
	\bus(16)(4,4);
	\bus(17)(2,4);
	\bus(18)(1,4);
	\bus(19)(5,4);
	\bus(20)(5,3);
	\bus(21)(4,5);
	\bus(22)(4,6);
	\bus(23)(3,6);
	\bus(24)(3,5);
	\bus(25)(0,4);
	\bus(26)(0,5);
	\bus(27)(2,5);
	\bus(28)(1,6);
	\bus(29)(0,6);
	\bus(30)(-1,3);
	\bus(31)(3,-1);
	\bus(32)(5,-1);
	\bus(33)(6,4);
	\bus(34)(6,3);
	\bus(35)(4,7);
	\bus(36)(3,7);
	\bus(37)(-1,4);
	\bus(38)(0,7);
	\bus(39)(0,1);
	\lin(1,0)(39,0);
	\lin(1,0)(2,0);
	\lin(2,-1)(30,1);
	\lin[-1](2,1)(3,-1);
	\lin(2,0)(25,0);
	\lin(4,-1)(3,1);
	\lin(3,-1)(18,-1);
	\lin(4,1)(14,-1);
	\lin(4,0)(5,0);
	\lin(5,-1)(8,1);
	\lin(5,0)(6,0);
	\lin(6,0)(31,0);
	\lin[-1](6,1)(11,-1);
	\lin[-1](6,-1)(7,1);
	\lin(7,1)(8,1);
	\lin(8,-1)(9,1);
	\lin(9,-1)(39,1);
	\lin(10,0)(32,0);
	\lin(10,0)(13,0);
	\lin[-1](10,-1)(11,1);
	\lin(11,-1)(12,-1);
	\lin[-1](12,1)(13,-1);
	\lin(13,-1)(14,1);
	\lin(14,0)(15,0);
	\lin(15,0)(16,0);
	\lin(16,0)(21,0);
	\lin[-1](16,-1)(17,1);
	\lin[-1](16,1)(19,-1);
	\lin(16,-1)(24,1);
	\lin[-1](17,-1)(18,1);
	\lin(17,1)(27,1);
	\lin(19,1)(33,-1);
	\lin(19,1)(20,1);
	\lin[-1](20,1)(34,-1);
	\lin(21,0)(22,0);
	\lin(22,0)(35,0);
	\lin[-1](22,-1)(23,1);
	\lin(23,0)(36,0);
	\lin(23,0)(24,0);
	\lin(25,0)(26,0);
	\lin(25,-1)(37,1);
	\lin[-1](26,1)(27,-1);
	\lin(26,0)(29,0);
	\lin(26,1)(28,-1);
	\lin(28,-1)(29,1);
	\lin(29,0)(38,0);
	\gen(1)(30,-.5);
	\gen(2)(31,-.5);
	\gen(3)(32,-.5);
	\gen(4)(33,1);
	\gen(5)(34,-1);
	\gen(6)(35,.5);
	\gen(7)(36,.5);
	\gen(8)(37,1);
	\gen(9)(38,.5);
	\gen(10)(39,-.5);
	\powin(1)(south,1);
	\powin(2)(west,.5);
	\powin(3)(west,.5);
	\powin(4)(north,1);
	\powin(5)(south,1);
	\powin(6)(east,.5);
	\powin(7)(west,.5);
	\powin(8)(north,1);
	\powin(9)(east,.5);
	\powin(10)(south,1);
	\load(3,0)(-1);
	\load(4,1)(-1);
	\load(5,1)(-1);
	\load(7,0)(-1);
	\load(8,0)(1);
	\load(12,0)(1);
	\load(15,1)(-1);
	\load(16,1)(1);
	\load(18,0)(-1);
	\load(20,0)(-1);
	\load(21,1)(1);
	\load(23,-1)(1);
	\load(24,-1)(-1);
	\load(25,1)(-1);
	\load(26,-1)(1);
	\load(27,0)(1);
	\load(28,0)(1);
	\load(29,-1)(1);
\end{tikzpicture}
\caption{Schematic of the 10-generator, 39-bus New England power grid model.}
\label{fig:case39-schematic}
\end{figure}

We also consider the 39-bus New England system~\cite{Pai1989}, shown in Fig.~\ref{fig:case39-schematic}, which is fed by 10 synchronous generators, and the swing generator $G_{10}$ models the interconnection to the external power grid.
The dynamics of the generator states are modeled with the Power Systems Toolbox (PST) \cite{Chow92, Pst92}.
New England system parameters can be found distributed with the software.
The software was modified to simulate noisy power injections modeled by K-L expansions and exponential autocovariance.

The mean mechanical power injections driving these generators are summarized in Table \ref{tab:pm2}. 
Autocorrelated-in-time noise in the power injections was introduced to generators $G_1$, $G_2$, and $G_3$, using a K-L expansion (\ref{eq:kl-Pm}) with $d_k = 25$ terms and covariance kernel $C_k(t,s) = \sigma^2\exp\left\{ -\left| t-s \right|/\lambda \right\}$, with standard deviation $\sigma = 0.02\text{ [p.u.]}$, and correlation length $\lambda = 1.8\text{ s}$. Then, at all points in time, the noise was transformed to have a log-normal stationary distribution but maintain the same mean and standard deviation.

Starting from deterministic equilibrium initial conditions, the system was solved for $20\text{ s}$, and at $10\text{ s}$ the transmission line between Bus 3 and Bus 4 experiences a 3-phase fault. This fault is cleared first at Bus 3 after $0.186\text{ s}$, and then at Bus 4 another $0.030\text{ s}$ later, after which system continues with the line removed. Note that for the same system with no input power noise, the critical clearing time (CCT) for Bus 3  is $0.189\text{ s}$, with the same lag in clearing time at Bus 4.

\begin{table} \setlength{\tabcolsep}{1ex}
  \centering
  \caption{Mean mechanical power injections [p.u.] for generator $G_j$.}
  \begin{tabular}{ccccccccccc}
    \hline\hline
      j & 1 & 2 & 3 & 4 & 5 & 6 & 7 & 8 & 9 & 10 \\
    \hline
     $\langle P_j^m\rangle$ & 0.25 & 0.60 & 0.65 & 0.63 & 0.1 & 0.68 & 0.56 & 0.54 & 0.83 & 1.01 \\
    \hline\hline
  \end{tabular}
  \label{tab:pm2}
\end{table}

\tpgset{BusConnectionSpacing}{.06in}	
\tpgset{BusLabelOffset}{0ex}		
\tpgset{BusLength}{0.21in}		
\tpgset{BusWidth}{2pt}			
\tpgset{LineTurnOffset}{0.4}		
\tpgset{LoadArrowLength}{0.6}		
\tpgset{GeneratorOffset}{1}		
\tpgset{InputPowerArrowLength}{1}	

\begin{figure}[h]
  \centering
  \includegraphics[width=0.36\textwidth]{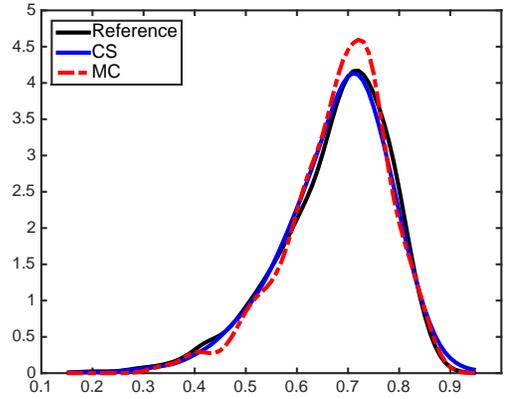}
  \caption{10-generator New England system: Comparison of the PDF of the voltage of bus 9 at time $t_{\mathrm{max}}$. ``MC'' indicate kernel density estimate computed from $500$ MC simulations; ``CS'' indicates kernel density estimate based on $10^4$ samples of $u^\ast$, which is constructed from the $500$ MC simulations; ``Reference'' indicates the reference kernel density estimate computed from $10^4$ MC simulations.}
  \label{fig:ex2_vol}
\end{figure}

\begin{figure}[h]
  \centering
  \includegraphics[width=0.36\textwidth]{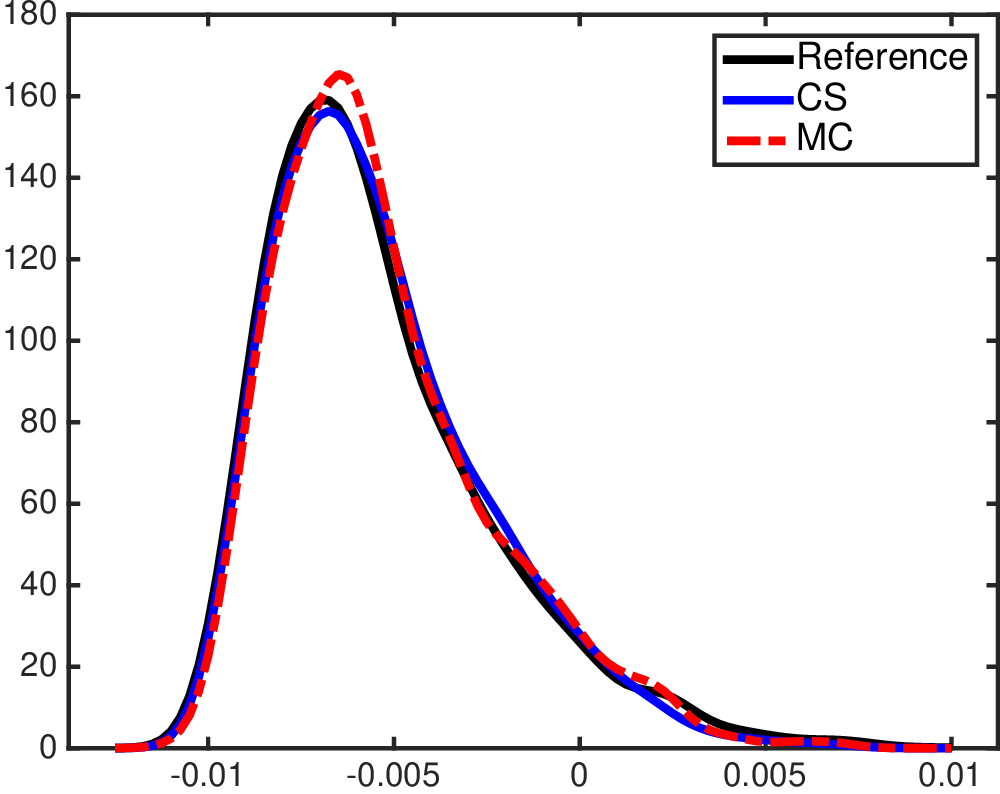}
  \caption{10-generator New England system: Comparison of the PDF of the relative angular velocity $\omega^\ast$ at time $t_{\mathrm{max}}$.
    ``MC'' indicate kernel density estimate computed from $500$ MC simulations; ``CS'' indicates kernel density estimate based on $10^4$ samples of $u^\ast$, which is constructed from the $500$ MC simulations; ``Reference'' indicates the reference kernel density estimate computed from $10^4$ MC simulations.}
  \label{fig:ex2_spd}
\end{figure}

PDF estimation for two QoIs is considered for this system: The voltage at the $9$th bus, and the difference of angular velocity between two specific generators $\omega^\ast(t_{\mathrm{max}})=\omega_{10}(t_{\mathrm{max}}) - \omega_{9}(t_{\mathrm{max}})$, at $t_{\mathrm{max}} = 13.51\text{ s}$.
Similar to the example in Section~\ref{sec:case9}, for each QoI $M = 500$ MC samples were used to construct $u^\ast$ with $d^\ast = 10, P^\ast = 4$, after which $u^\ast$ was sampled $10^4$ times to estimate the PDF via kernel density estimation.
A reference MC PDF estimate is computed via kernel density estimation directly from $10^4$ samples of the full system.
Additionally, a direct kernel density estimate was produced from the $M = 500$ samples employed to construct the surrogate models.
The MC estimates and the PDF estimate from $u^\ast$ are shown for both QoIs in Figures~\ref{fig:ex2_vol} and~\ref{fig:ex2_spd}.

KL divergences of the PDF estimates with respect to the reference PDF for both quantities of interest are presented in Tables~\ref{tab:ex2_vol} and \ref{tab:ex2_spd}.
As in the numerical example of Section~\ref{sec:case9}, the KL divergences for PDF estimates were obtained using $50$ independent sample sets, each set consisting of $500$ samples.
Again, the sparse regression estimate of the PDF proved to be more accurate than the direct kernel density estimate obtained from the same number of samples as were used to construct the surrogate model.
Furthermore, the difference between PDF estimates obtained from the $\tilde u$ (rotation without truncation) and $u^\ast$ (truncation after rotation) models was small.

\begin{figure}[h]
  \centering
  \includegraphics[width=0.36\textwidth]{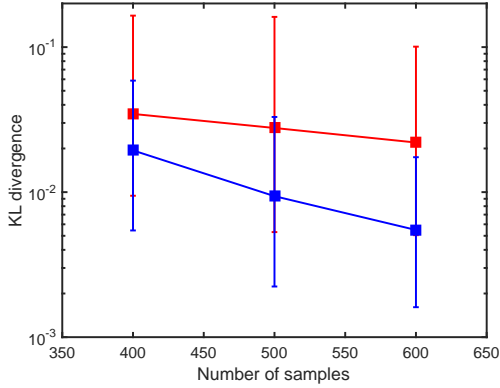}
  \caption{10-generator New England system: Comparison of mean KL divergence of approximations to the PDF of the voltage of bus 9 at time $t_{\mathrm{max}}$ with respect to the reference density.
    Red bars indicate direct kernel density estimate of PDF from MC simulations; blue bars indicate kernel density estimate of PDF from samples of $u^\ast$.}
  \label{fig:ex2_vol_kl}
\end{figure}

\begin{figure}[h]
  \centering
  \includegraphics[width=0.36\textwidth]{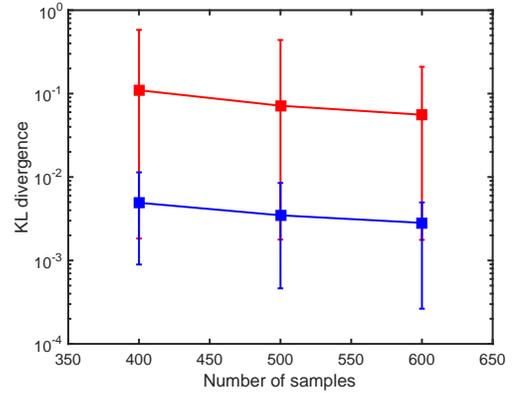}
  \caption{
    10-generator New England system: Comparison of mean KL divergence of approximations to the PDF of the relative angular velocity $\omega^\ast$ at time $t_{\mathrm{max}}$ with respect to the reference density.
    Red bars indicate direct kernel density estimate of PDF from MC simulations; blue bars indicate kernel density estimate of PDF from samples of $u^\ast$.}
  \label{fig:ex2_spd_kl}
\end{figure}

Also studied were the effects of varied data availability for estimating the PDF for both QoIs. $50$ independent datasets were employed to study the statistics of the KL divergence of the various approximations to the PDF.
Figures~\ref{fig:ex2_vol_kl} and~\ref{fig:ex2_spd_kl} show the mean KL divergence together with its $99\%$ confidence interval indicated by error bars.
It is evident that the sparse reduction method again improves upon density estimation from MC samples, in terms of consistently lower KL divergence.
\begin{table}
  \centering
  \caption{10-generator New England system: Mean KL divergence of various approximations to the PDF of the voltage of bus 9 at time $t_{\mathrm{max}}$, computed using different methods (with $500$ samples), with respect to the reference density.}
  \begin{tabular}{C{8em}C{8em}C{8em}}
    \hline\hline
       MC & $\tilde u$  & $u^\ast$ \\
    \hline
    $0.0375$ & $0.0142 $  & $0.0098$ \\
    \hline\hline
  \end{tabular}
  \label{tab:ex2_vol}
\end{table}

\begin{table}
  \centering
  \caption{10-generator New England system: Mean KL divergence of various approximations to the PDF of the relative angular velocity $\omega^\ast$ at time $t_{\mathrm{max}}$, computed using different methods (with $500$ samples), with respect to the reference density.}
  \begin{tabular}{C{8em}C{8em}C{8em}}
    \hline\hline
       MC & $\tilde u$  & $u^\ast$ \\
    \hline
    $0.0746 $ & $0.0032 $  & $0.0029$ \\
    \hline\hline
  \end{tabular}
  \label{tab:ex2_spd}
\end{table}

\section{Conclusions}
\label{sec:conclusions}
Due to the nonlinear nature of power grid dynamics, the distribution of QoIs subject to uncertain inputs is non-Gaussian and may exhibit long tail behavior; therefore, central moments do not sufficiently characterize such distributions for the purposes of risk assessment and optimization under uncertainty.
Therefore, accurate estimation of PDFs is a critical necessity for the analysis and operation of power grid systems under uncertainty.

In this study, surrogate models were constructed for QoIs of power systems subject to random power injection fluctuations.
A compressive sensing method based on iterative rotations was used to construct the surrogate model, which requires a relatively small number of samples from the full model.
The PDF of the QoI is then estimated by sampling the surrogate model.

The proposed numerical method was applied to two power grid systems.
Sparse regression was shown to be more efficient than the traditional MC method. 
KL divergence was used to compute the error in PDF estimate relative to the reference solution and show that for the same number of samples of the full system, sparse regression exhibited KL divergences approximately $75\%$ smaller than those for the MC method.

\section*{Acknowledgments}
\label{sec:acknowledgments}

This work was supported by the Applied Mathematics Program within the U.S. Department of Energy Office of Advanced Scientific Computing Research as part of the Multifaceted Mathematics for Complex Systems project.
Pacific Northwest National Laboratory is operated by Battelle for the DOE under Contract DE-AC05-76RL01830.
X. Yang and D.~A. Barajas-Solano contributed equally to this manuscript.

\bibliographystyle{IEEEtran}
\bibliography{IEEEabrv,power-pdf-est}

\begin{IEEEbiographynophoto}{Xiu Yang}
  received his B.Sc. and M.Sc. from Peking University, Beijing, China, and Ph.D.
  from Brown University in 2005, 2008, and 2014, respectively.
  He is currently a scientist at the Computational Mathematics group at the Pacific Northwest National Laboratory, Richland, Washington, USA.
  His research interests include uncertainty quantification, data assimilation, inverse problem and multiscale modeling.
  He can be reached at \emph{xiu.yang@pnnl.gov}.
\end{IEEEbiographynophoto}
\begin{IEEEbiographynophoto}{David A. Barajas-Solano}
  received his B.Sc. from the Industrial University of Santander, Bucaramanga, Colombia, and M.Sc. and Ph.D. from the University of California, San Diego, in 2008, 2010, and 2013, respectively.
  He is currently a scientist at the Computational Mathematics group at the Pacific Northwest National Laboratory, Richland, Washington, USA.
  His research interests include stochastic differential equations and DAEs, PDF methods for uncertainty quantification, and multiscale modeling.
  He can be reached at \emph{David.Barajas-Solano@pnnl.gov}.
\end{IEEEbiographynophoto}
\begin{IEEEbiographynophoto}{W. Steven Rosenthal}
  received his B.Sc. in Mechanical Engineering from Arizona State University in 2008, and Masters and Ph.D. degrees in Applied Mathematics from the University of Arizona in 2010 and 2014, respectively.
  He is currently a postdoctoral research assistant in the Computational Mathematics group at Pacific Northwest National Laboratory in Richland, WA.
  His research interests include numerical analysis and methods for solving PDEs/SDEs/SDAEs, and uncertainty quantification including inverse modeling and data assimilation.
  He can be reached at \emph{william.rosenthal@pnnl.gov}.
\end{IEEEbiographynophoto}
\begin{IEEEbiographynophoto}{Alexandre M. Tartakovsky}
  is the Associate Division Director for Computational Mathematics in the Pacific Northwest National Laboratory's Advanced Computing, Mathematics, and Data Division.
  His research focuses on multiscale mathematics and uncertainty quantification with application to complex natural and engineered systems.
  Dr.~Tartakovsky has received 2011 DOE Early Career award and was recognized with a Presidential Early Career Award for Scientists and Engineers in 2009, for his research on subsurface flow that addresses past and future energy needs.
  He earned a Master¹s degree in hydromechanics and applied mathematics from Kazan State University in Russia in 1994, and a Ph.D. in hydrology from the University of Arizona in Tucson in 2002.
  He has joined PNNL in 2004 after a two-year postdoctoral appointment at DOE's Idaho National Laboratory.
  He can be reached at \emph{alexandre.tartakovsky@pnnl.gov}.
\end{IEEEbiographynophoto}

\end{document}